\documentclass[10pt, pra, floatfix,twocolumn]{revtex4}

\usepackage{hyperref}
\usepackage{enumerate}
\usepackage{ifthen}

\usepackage{verbatim}
\usepackage[usenames,dvipsnames]{pstricks}
\usepackage{pst-node}   
\usepackage{epsfig}
\usepackage{amsmath}
\usepackage{latexsym}
\usepackage{revsymb}
\usepackage{yfonts}

\usepackage{natbib}
\usepackage{amsfonts}
\usepackage{amsmath}
\usepackage{amssymb}
\usepackage{amsthm}
\usepackage{graphicx}
\usepackage{bm}
\usepackage{bbm}

\newcommand{\be}{\begin{eqnarray} \begin{aligned}}
\newcommand{\ee}{\end{aligned} \end{eqnarray} }
\newcommand{\benn}{\begin{eqnarray*} \begin{aligned}}
\newcommand{\eenn}{\end{aligned} \end{eqnarray*} }

\newcommand{\Mat}{\mathbb{M}}
\newcommand*{\Enc}{\mathsf{Enc}}
\newcommand*{\Ext}{\mathsf{Ext}}

\newcommand{\bc}{\begin{center}}
\newcommand{\ec}{\end{center}}

\newcommand{\id}{\mathbb{I}}

\newcommand{\tr}{\mathop{\mathrm{tr}}\nolimits}

\newtheorem{theorem}{Theorem}

\newtheorem{lemma}{Lemma}
\newtheorem{definition}{Definition}

\newcommand{\hmin}{\ensuremath{{\rm H}_{\infty}}}
\newcommand{\hmineps}{\hmin^{\varepsilon}}

\usepackage{amsfonts}

\def\id{\mathbb{I}}

\newcommand*{\sbin}{\{0,1\}}

\newcommand{\eps}{\varepsilon}
\newcommand{\ket}[1]{|#1\rangle}
\newcommand{\bra}[1]{\langle#1|}

\newcommand{\proj}[1]{|#1\rangle\langle#1|}

\newcommand{\inp}[2]{\langle{#1}|{#2}\rangle} 

\newcommand*{\cA}{\mathcal{A}} 
\newcommand*{\cB}{\mathcal{B}}

\newcommand*{\cE}{\mathcal{E}}
\newcommand*{\cF}{\mathcal{F}}

\newcommand*{\cH}{\mathcal{H}}
\newcommand*{\cI}{\mathcal{I}}

\newcommand*{\cN}{\mathcal{N}}

\newcommand*{\cR}{\mathcal{R}}

\newcommand*{\cS}{\mathcal{S}}

\newcommand*{\cT}{\mathcal{T}}

\newcommand{\hout}{\mathcal{H}_{\rm out}}
\newcommand{\hin}{\mathcal{H}_{\rm in}}

\newcommand{\scp}{strong-converse property}

\newcommand{\bop}{{\mathcal{B}}}

\newcounter{protoCount}
\newcounter{protoList}
\newsavebox{\tmpbox}

\newenvironment{protocol}[3]{
\bigskip
\addtocounter{protoCount}{1}
\noindent
\newline
\begin{bfseries}Protocol #1: #2\end{bfseries}
\ifthenelse{\equal{#3}{\empty}}{}{\\ #3}
\begin{list}{\begin{bfseries}\arabic{protoList}:\end{bfseries}}
{\usecounter{protoList}}
}{
\end{list}
\bigskip
}

\newcommand{\assign}{:=}

\begin{document}

\title{Achieving the physical limits of the bounded-storage model}
\author{Prabha Mandayam}
\affiliation{Institute for Quantum Information, California Institute of Technology, Pasadena CA 91125, USA}
\author{Stephanie Wehner}
\affiliation{Centre for Quantum Technologies, National University of Singapore, 2 Science Drive 3, 117543 Singapore}
\date{\today}

\begin{abstract}
Secure two-party cryptography is possible if the adversary's quantum storage device suffers imperfections.
For example, security can be achieved if the adversary
can store strictly less then half of the qubits transmitted during the
protocol. This special case is known as the bounded-storage model, and it has long been an open question
whether security can still be achieved if the adversary's storage were any larger.
Here, we answer this question positively and demonstrate a two-party protocol which is secure
as long as the adversary cannot store even a small fraction of the transmitted pulses. We also show that security can be extended to a larger class of noisy quantum memories.
\end{abstract}  

\maketitle

\section{Introduction}

Two-party cryptography enables Alice and Bob to solve problems in cooperation even if they do not trust each other.
Important examples of such tasks include auctions and secure identification. 
In the latter, Alice wants to identify herself to Bob (possibly a fraudulent ATM machine) without revealing her password.
\begin{figure}
\begin{center}
\label{fig:secureid}
\includegraphics{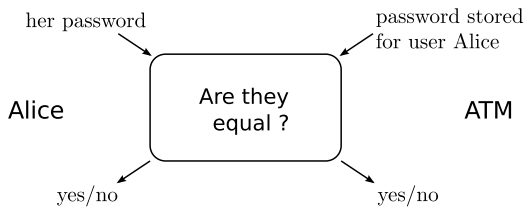}
\end{center}
\caption{Secure Identification}
\end{figure}
More generally, Alice and Bob wish to solve problems where Alice holds an input $x$ (e.g. her password)
and Bob holds an input $y$ (e.g. the password an honest Alice should possess), and 
they want to obtain the value of some function $f(x,y)$ (e.g. `yes' if $x = y$, and `no' otherwise), as depicted in Fig. \ref{fig:secureid}. Security means that Alice should not learn anything about $y$ and Bob should not learn anything about $x$, apart from what can be inferred from $f(x,y)$~\cite{yao:sfe}. 

Contrary to quantum key distribution where honest Alice and Bob can work together to detect the presence of
an outside eavesdropper~\cite{bb84,e91}, two-party cryptography is made difficult by the fact
that Alice and Bob do not trust each other and have to fend for themselves.
Indeed, two-party cryptography is impossible without making assumptions about the adversary, even when we allow 
quantum communication~\cite{lo:insecurity}. 
The security of most cryptographic systems in use today is based on the premise 
that certain computational problems are hard to solve for the adversary. Concretely, security 
relies on the assumption that the adversary's computational resources are limited, 
and the underlying problem is hard in some precise complexity-theoretic sense. While the former assumption may be justified in practice, 
the latter statement is usually an unproven mathematical conjecture.

It is thus a natural question whether other, more physical assumptions regarding the two parties' resources allow us to obtain security without
relying on any additional hardness results. This is indeed known to be possible if we assume that
the adversary's classical~\cite{maurer1,maurer2,cachin:boundedOld} or quantum storage
is limited~\cite{serge:bounded,serge:new,bounded:secureId} 
or more generally if his memory is simply imperfect~\cite{prl:noisy,noisy:robust,kww:nstorage}. 

Concretely, the assumption of the noisy-storage model entails that during waiting times $\Delta t$ in a protocol, the adversary
has to measure/discard all his quantum information except what he can encode (arbitrarily) into his quantum memory.
Any quantum storage can be modeled as a completely positive trace preserving (CPTP) map $\cF: \bop(\hin) \rightarrow \bop(\hout)$,
that maps states $\rho \in \hin$ to some noisy states $\cF(\rho) \in \hout$. In this paper, we focus on the case where the input space is an $n$-fold tensor product $\hin\cong(\mathbb{C}^d)^{\otimes \nu \cdot n}$, the protocols involve $n$-qu$d$its of communication and the noise is of the form
$\cF\equiv \cN^{\otimes \nu \cdot n}$ with
$\cN:\cB(\mathbb{C}^d)\rightarrow\cB(\mathbb{C}^d)$. The constant $\nu > 0$ is referred to as
the \emph{storage rate} as it captures the fraction of the transmitted qu$d$its that could 
could potentially be stored by the adversary.

Clearly, the storage rate $\nu$ plays a crucial role in deciding whether security can be obtained from a particular storage device. For example, in the case of bounded storage where we have no noise ($\cN = \id$),
we can never hope to obtain security if the adversary can store all quantum information made available
to him during the protocol, that is, if $\nu = 1$ and the input space is $\hin = (\mathbb{C}^d)^{\otimes n}$. Apart from this trivial condition, however, no bounds were known that restrict our ability to obtain security. In~\cite{serge:new} it was shown that security can be
achieved in a protocol based on qubits ($d=2$) as long as $\nu < 1/4$. 
This was improved to $\nu < 1/2$ in~\cite{kww:nstorage}. More generally, it was shown that
security in the noisy-storage model can be obtained~\cite{kww:nstorage} if
\begin{align}\label{eq:originalBound}
C_{\mathcal{N}} \cdot \nu < \frac{1}{2}\ ,
\end{align}
where $C_{\mathcal{N}}$ is the classical capacity of the quantum channel $\mathcal{N}$. 

\subsection{Results}
Here, we show that for the case of bounded storage, security can be obtained if the cheating party can store all but a constant fraction 
of the transmitted pulses. That is, the trivial condition $\nu < 1$ stated above is in fact optimal! 
The honest players thereby need no quantum storage at all in order to execute the protocol.
This not only settles the question, but also highlights the sharp contrast to the case of classical bounded
storage, where it was shown that security can only be obtained if the adversary's classical 
storage is at most quadratic in the storage required by the honest players~\cite{maurer:imposs}.
Unlike the protocols in~\cite{serge:bounded,serge:new, prl:noisy, kww:nstorage} which use
BB84 encoded qubits~\cite{bb84}, we make use of states encoded in higher-dimensional mutually unbiased 
bases~\footnote{Two orthonormal bases 
$\mathcal{B}_1 = \{\ket{x^{(1)}} \mid x \in \{0,\ldots,d-1\}\}$ and
$\mathcal{B}_2 = \{\ket{x^{(2)}} \mid x \in \{0,\ldots,d-1\}\}$ are called mutually unbiased if for all $x, y \in \{0,\ldots,d-1\}$
we have $|\inp{x^{(1)}}{y^{(2)}}| = 1/\sqrt{d}$.}. Of course, we also scale the storage size 
accordingly to
$\hin = (\mathbb{C}^d)^{\otimes \nu \cdot n}$ when sending $d$ dimensional states. More specifically, we show that security in the setting of bounded
storage is possible as long as
\begin{align}\label{eq:boundedClaim}
	\nu < \frac{\log(d+1) -1}{\log d} \rightarrow 1\ ,
\end{align}
where the r.h.s. approaches $1$ for large $d$. We stress that for large values of $d$, the resulting protocols
will be much harder to implement experimentally, and even though the errors decrease exponentially with $n$ 
they converge very slowly for large $d$. Note, however, that here we are merely interested in exploring the fundamental physical limitations of this model. 

For the general setting of noisy quantum storage we further show that security is possible for devices
$\cF = \cN^{\otimes \nu \cdot n}$, where the channel $\cN: \bop(\mathbb{C}^d) \rightarrow \bop(\mathbb{C}^d)$ satisfies the strong converse property~\cite{rs:converse}, whenever
\begin{align}
	C_{\mathcal{N}} \cdot \nu < \log (d+1) - 1\  ,
\end{align}
thus extending the range of storage devices for which we can prove security~\cite{kww:nstorage}.
Our proof thereby relies on an uncertainty relation for mutually unbiased bases, but is completely general in the sense that
any other set of encodings satisfying such a relation could be used in our protocol instead.

We would like to emphasize that that the setting considered here differs greatly from
quantum key distribution (QKD)~\cite{bb84,e91}, where higher dimensional states have also been used to some advantage. 
In QKD, Alice and Bob \emph{trust each other}, but are trying to protect themselves
from an outside eavesdropper. An important advantage gained by Alice and Bob in this setting is that they can work together to try and
\emph{detect} interference by such an eavesdropper. In contrast, in the scenario we are considering there is no analogous way for Alice to check
on any of Bob's actions, and vice versa. Hence, we require an entirely different proof of security as used in quantum key distribution, and
whereas results from QKD may provide some clues, they give only very little indication that higher dimensional states are useful for our problem.

\subsection{Techniques}
We first give an overview of the steps involved in obtaining our result.
The constant~$1/2$ in the bound~\eqref{eq:originalBound} is a result of using BB84 states~\cite{bb84} in the protocol,
and stems from an uncertainty relation for measurements in these two bases~\cite{maassen:entropy}. 
It is thus natural to consider a protocol that uses more than two mutually unbiased bases (MUBs) for which
uncertainty relations are known to exist~\cite{sanchez:entropy}.
Our first step is to obtain a modified protocol for the simple two-party primitive weak string erasure which was originally introduced in~\cite{kww:nstorage}, using the full set of $d+1$ MUBs that are known to exist in prime power dimensions~\cite{boykin:mub,wootters:mub}. 
Next, we show that there is still a secure protocol for the cryptographic primitive of oblivious transfer using this variant of weak string erasure. This is done by purely classical post-processing of the output of the quantum primitive weak string erasure. Since it is known that any two-party cryptographic problem can be solved using oblivious transfer~\cite{kilian:foundingOnOT}, this concludes our claim. 

\section{Weak String Erasure}

We first show how to obtain a variant of the very simple cryptographic primitive weak string erasure (WSE) introduced in~\cite{kww:nstorage}, which we will call \emph{non-uniform} weak string erasure;
a formal definition can be found in Appendix~\ref{sec:wse}.
Intuitively, this primitive provides Alice with a 
string $X^{n} = (X_{1},...,X_{n}) \in \{0,1,\ldots,d-1\}^{n}$, where each entry $X_{i}$ takes on one of $d$ possible values. 
Bob obtains a set of index locations $\mathcal{I} = \{i_{1},...,i_{|\mathcal{I}|}\mid i_j \in [n]\}$, where any index $i \in \{1,\ldots,n\} =: [n]$ is chosen to be in $\cI$ with some probability $p$.
In addition, Bob receives the entries of the string $X^n$ corresponding to the indices $\cI$, 
which we denote by the substring $X_{\mathcal{I}} = (X_{i_{1}},X_{i_{2}},\ldots,X_{i_{|\cI|}})$. 
Security here means that even if Alice is dishonest, she cannot learn which entries are known to Bob, i.e., she cannot
learn anything about the index set $\cI$. Conversely, if Bob is dishonest, then his information about the entire string $X^n$ should still be limited in the sense that
the probability that he can guess all of $X^n$ 
given his information $B'$ is small. That is,
\begin{align}
	P_{\rm guess}(X|B') \leq 2^{- \lambda n}\ .
\end{align}
for some $\lambda > 0$.
This is equivalent~\cite{krs:entropy}
to demanding that his min-entropy~\cite{renato:diss} denoted as $\hmin(X^{n}|B')_\rho$, obeys
\begin{align}
\hmin(X^{n}|B')_\rho=-\log P_{guess}(X^{n}|B') \geq \lambda n\ . \label{eq:minentropyguess}
\end{align}
In practice, we allow this condition to fail with error parameter $\eps$, which corresponds to demanding that the \emph{smooth} min-entropy defined as
\begin{align}
&\hmineps(X^n|B')_\rho \hspace{-0.5mm}\\
&\qquad :=\sup_{\substack{
\bar{\rho}_{X^nB'} \geq 0\\
\frac{1}{2}\|\bar{\rho}_{X^nB'}-\rho_{X^nB'}\|_1\leq \tr(\rho_{X^nB'})\cdot \varepsilon\\
\tr(\bar{\rho}_{X^nB'})\leq \tr(\rho_{X^nB'})
 }}\hspace{-0.7mm}\hmin(X^n|B')_{\bar{\rho}}\ ,\nonumber
\end{align}
satisfies 
\begin{align}\label{eq:BobSecurity}
	\hmineps(X^n|B') \geq \lambda n\ .
\end{align}

\subsection{Protocol}

Next we outline a simple protocol that achieves the functionality described above. It is a straightforward generalization of the original protocol in~\cite{kww:nstorage} to multiple encodings, the main difference being that the indices in $\cI \subseteq [n]$ are no longer chosen uniformly at random. Instead, the probability $p$ that honest Bob learns the value of $X_i$ for $i \in [n]$ is equal to the probability that 
he chooses the same basis as Alice, that is, $p = 1/(d+1)$. In what follows, let $2^{[n]}$ denote the set of all subsets of $[n]$.

\begin{protocol}{1}{Non-uniform WSE}{Outputs: $x^n \in \{0,1,\ldots,d-1\}^n$ to Alice, $(\cI,z^{|\cI|}) \in 2^{[n]} \times \{0,1,\ldots,d-1\}^{|\cI|}$ to Bob.}\label{proto:wse}
\item {\bf Alice: } Picks an $n$-$d$it string uniformly at random, $x^{n} \in \{0,1,...,d-1\}^{n}$. She encodes each $d$it into one of the $d+1$ MUBs, $\cB_{\theta_1},\ldots,\cB_{\theta_n}$, that is, she chooses a  basis string $\theta^{n} = (\theta_1,\ldots,\theta_n) \in \{0,...,d\}^{n}$ uniformly at random, so that the $d$it $x_{j}$ is encoded in basis $\cB_{\theta_{j}}$, and sends it to Bob.
\item {\bf Bob: } Chooses a basis string $\tilde{\theta}^{n} \in \{0,1,...,d\}^{n}$ uniformly at random. When receiving the $i$-th qu$d$it, he measures it in the basis $\cB_{\tilde{\theta}_{i}}$, to obtain outcome $\tilde{x}_{i}$.
\item[Both parties wait time $\Delta t$.]
\item {\bf Alice: } Sends the basis information $\theta^{n}$ to Bob, and outputs $x^{n}$.
\item {\bf Bob: } Computes $\cI := \{ i \in [n] | \theta_{i} = \tilde{\theta}_{i} \}$, and outputs $(\cI,\tilde{x}_{\cI})$.
\end{protocol}

We now formally state our claim that this protocol does indeed implement non-uniform WSE, with the parameters $\eps$, $\lambda$ and $d$.
\begin{theorem}\label{thm:mainclaimwse}
Let Bob's storage be given by $\cF = \cN^{\otimes \nu n}$ with a storage rate~$\nu>0$, where $\cN$ satisfies the strong converse property~\cite{rs:converse} and the capacity $C_\cN$ of the channel $\cN$ bounded by
\begin{align}
C_\cN\cdot\nu<\log(d+1)-1\ . \label{eq:capacity}
\end{align} 
Let $\delta\in ]0,\frac{1}{2}-C_\cN\cdot \nu[$. Then, Protocol~1 securely implements weak string erasure for sufficiently 
large $n$ with
\begin{align}
\lambda(\delta,d) &= \nu \cdot \gamma^\cN\left(\frac{\log(d+1)-1-\delta}{\nu}\right)\, \label{eq:min-entropy}
\end{align}
and error $\eps(\delta,d) = 2\exp(-f(\delta,d)n)$ with 
\begin{equation}
f(\delta,d) \propto -\delta^2/\left(\log((d+1)\cdot d)+ \log 4/\delta\right)^2 > 0,
\end{equation}
where $\gamma^{\cN}(\cdot)$ is the strong converse parameter of $\cN$~\cite{rs:converse}.
\end{theorem}

Note that for bounded storage, where $\mathcal{N}$ is simply the identity channel over Bob's $d$-dimensional input Hilbert space, $C_{\mathcal{N}} = \log d$ in~\eqref{eq:capacity}, which directly implies our central result~\eqref{eq:boundedClaim}.

It is easy to see that the protocol is correct if both parties are honest: if Alice is honest, her string $X^{n} = x^{n}$ is chosen uniformly at random from $\{0,1,\ldots,d-1\}^{n}$ as desired, and if Bob is honest, he clearly obtains $\tilde{x}_{i} = x_{i}$ whenever $i\in \cI$ for a random subset $\cI \subseteq [n]$. In the remainder of this section, we prove security if one of the parties is dishonest.

\subsection{Security against dishonest Bob}
First of all, we need to show that even if Bob is dishonest, he can nevertheless not learn much about the entire string $X^n$. In other words, our goal is to show
that there exists some $\lambda > 0$ satisfying~\eqref{eq:BobSecurity}.
Our proof proceeds in three steps; technical details can be found in Appendix~\ref{sec:wse}. First, we consider Bob's information about the string $X^n$ given only his classical information $K$,
and the basis information $\Theta^n$ he receives. This can be quantified using entropic uncertainty relations in terms of the Shannon entropy for $d+1$ MUBs 
in $\mathbb{C}_{d}$~\cite{sanchez:entropy}.

Using~\cite[Theorem 4.22]{chris:diss} these uncertainty relations imply a bound on Bob's information in terms of the smooth min-entropy
\begin{equation}\label{eq:classicalUR}
H^{\varepsilon/2}_{\infty}(X^{n}|K\Theta^{n})_{\rho} \geq \left(\log(d+1)-1 - \frac{\delta}{2}\right)n\ ,
\end{equation}
for any $0<\delta<\frac{1}{2}$ with $\varepsilon = 2\exp\left(-f(\delta,d) n\right)$, for $f(\delta,d) \geq 0$. That is, the error decreases exponentially with $n$, as desired.
Note that instead of mutually unbiased bases, we could have used any other form of encodings obeying such a strong uncertainty relation.

\begin{figure}[ht]
\begin{center}
\includegraphics{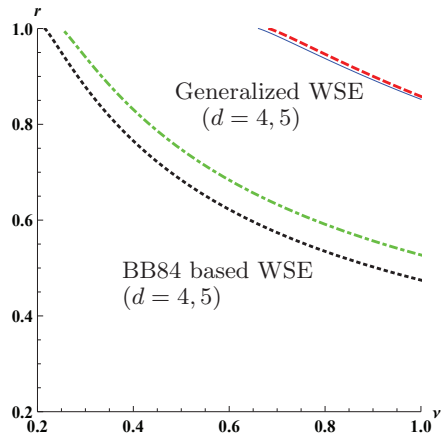}
\end{center}
\caption{Security regions $(r,\nu)$ for weak string erasure (WSE) with depolarizing noise $\mathcal{N}(\rho) = r \rho+ (1-r) \id/d$ , in dimensions $d=4,5$. Previously~\cite{kww:nstorage}, security was shown in the regions below the dotted black curve for $d=4$ and the dot-dashed green curve for $d=5$. Our analysis extends the security region to the solid blue curve ($d=4$) and the dashed red curve ($d=5$) respectively. 
\label{fig:comparison}}\end{figure}

Next we consider Bob's information, when in addition he is given the output of his storage device $\cF(Q)$. We know from~\cite{kww:nstorage} that the uncertainty relation~\eqref{eq:classicalUR} determines the rate at which Bob needs to send information through 
his storage device. Using~\cite[Lemma 2.2]{kww:nstorage} together with~\eqref{eq:classicalUR} we obtain
\begin{align}
& H^{\varepsilon}_{\infty}(X^{n}|\Theta^{n}K \cF(Q))_{\rho} \nonumber \\
&\qquad\geq -\log P^{\mathcal{F}}_{succ}\left[n\left(\log(d+1)-1-\frac{\delta}{2}\right)\right]\ ,\label{eq:rate}
\end{align}
where, $P^{\cF}_{\rm succ}(nR)$ is the average probability of sending a randomly chosen string $x \in \{0,1\}^{nR}$ through the storage $\cF$~\footnote{$P^{\cF}_{\rm succ}(nR)$ is given by $P^{\cF}_{\rm succ}(nR) = \max_{\{M_x\}_x,\{\rho_x\}_x} \frac{1}{2^{nR}} \sum_{x \in \{0,1\}^n} \tr\left(M_x \cF(\rho_x)\right)$, where the maximization is taken over encodings $\{\rho_x\}_x$ and decoding positive operator valued measures (POVMs) $\{M_x\}_x$.}.
For noise of the form $\cF = \cN^{\otimes \nu n}$, the r.h.s. of~\eqref{eq:rate} is the success probability of sending $\nu n$ bits at a rate $R = (\log(d+1) - 1 - \delta/2)/\nu$. The final step is to note that for channels satisfying
the strong converse property~\cite{rs:converse}, this success probability drops off exponentially with $n$ according 
to the parameter $\gamma^{\cN}(\cdot)$, whenever $R > C_{\mathcal{N}}$. This gives the bound
\begin{align}
	C_{\mathcal{N}} \cdot \nu < \log(d+1) - 1 - \frac{\delta}{2}\ .
\end{align}
Theorem~\ref{thm:mainclaimwse} then follows by noting that exponential security (in $n$) is possible for any constant $\delta > 0$. As an example of how our bound improves upon the earlier bound in~\cite{kww:nstorage}, we compare the corresponding security regions for WSE with depolarizing noise, i.e. when $\cN(\rho) = r\rho + (1 - r)\id/d$, in Fig.~\ref{fig:comparison}.

\subsection{Security against dishonest Alice}
When Alice is dishonest, it is intuitively obvious that she is unable to gain any information about the index set $\cI$, since she never receives any information from Bob during our protocol. However, a more careful security analysis is required if we want to use weak string erasure to build more complicated primitives like oblivious transfer. This argument is essentially analogous to~\cite{kww:nstorage}, as outlined in Appendix~\ref{sec:wse} for completeness. 

\section{Oblivious transfer}

Ultimately, we would like to use WSE to solve arbitrary two-party cryptographic problems. To this end, it suffices to implement the primitive oblivious transfer~\footnote{Here we execute fully randomized oblivious transfer~\cite{serge:new,fs:compose} which can easily be converted into standard 1-2 oblivious transfer~\cite{BBCS92,Beaver95}.}, which can solve any two-party problem~\cite{kilian:foundingOnOT}. Informally, this primitive outputs
two strings $S_0^\ell,S_1^\ell \in \{0,1\}^\ell$ to Alice, and a choice bit $C \in \{0,1\}$ and $S_C^\ell$ to Bob. Security means that if Alice is dishonest, she should not learn anything about $C$. If Bob is dishonest, we demand that there exists some random variable $C$ such that Bob is entirely ignorant about $S_{1-C}^\ell$. That is, he may learn at most one of the two strings which are generated. 

Here, we state a simplified version of the actual protocol which executes fully randomized oblivious transfer from WSE. This is a purely classical protocol,
using the quantum primitive WSE. 
It contains all the essential ingredients to understand the main steps of our security proof. A formal definition, as well as the full protocol can be found in Appendix~\ref{sec:OT}. The only difference from the protocol presented in~\cite{kww:nstorage} is the fact that $\cI$ is no longer uniform, and honest Bob
only learns about $pn$ entries $x_j$, whereas in the case of uniform WSE~\cite{kww:nstorage} he could learn roughly $n/2$. 
We hence introduce a new parameter $\eta = 2(d+1)$ in the protocol, such that with high probability Bob learns at least $n/\eta$ of the indices.

Our protocol uses two ingredients, privacy amplification and a primitive called interactive hashing, where we refer to~\cite{kww:nstorage}
for an overview of these techniques. Privacy amplification is well-known from its role in quantum key distribution~\cite{renato:diss}. Although interactive hashing is well-known within the realm of classical cryptography, it has only recently found applications in quantum information~\cite{kww:nstorage}.
Intuitively, an interactive hashing protocol has the following properties: It takes as inputs a subset $\cI_{\rm tr}$ (encoded as a string $w$)
from Bob, and outputs two subsets $\cI_0, \cI_1 \in [n]$ (encoded as strings $w_0,w_1$) to both Alice and Bob. The protocol ensures that there
exists a $c \in \{0,1\}$, such that $\cI_c = \cI_{\rm tr}$, i.e., one of the two subsets it outputs is equal to Bob's original input. Note that
since Bob knows his input, he can of course compute $c$. Nevertheless, interactive hashing ensures that Alice cannot learn which subset is the same as
Bob's input, that is, Alice cannot learn $c$. Finally, interactive hashing has another important property we will need: Whereas Bob can choose
one of these subsets (namely $\cI_c$), the choice of the other subset is not under his control. In fact, $\cI_{1-c}$ is essentially
chosen at random.

\begin{protocol}{2}{Oblivious Transfer}{Outputs: $(s_0^\ell,s_1^\ell) \in
\{0,1\}^{\ell} \times \{0,1\}^\ell$ to Alice, and $(c,y^\ell) \in \{0,1\} \times \{0,1\}^\ell$
to Bob}
\item {\bf Alice and Bob: } Execute WSE. Alice gets a string $x^n \in \{0,1, \ldots, d-1\}^n$, Bob a set $\cI \subset [n]$ and a string $s = x_\cI$. If $|\cI| < n/\eta$, Bob chooses uniformly at random a set $\cI_{\rm tr}$ of size $|\cI_{\rm tr}| = n/\eta$.
Otherwise, he
randomly truncates $\cI$ to $|\cI_{\rm tr}|$ of size~$n/\eta$, and deletes the
corresponding values in $s$.
\item {\bf Alice and Bob:} Execute interactive hashing with Bob's input~$w$ equal to a description of~$\cI_{\rm tr}=\Enc(w)$. 
Interpret the outputs $w_0$ and $w_1$ as descriptions of subsets $\cI_0$ and $\cI_1$ of~$[n]$.
\item {\bf Alice:} Chooses $r_{0}, r_{1} \in_R \cR$ and sends them to Bob.
\item {\bf Alice:} Outputs $(s_0^\ell,s_1^\ell)$~$\assign$~$(\Ext(x_{\cI_0},\hspace{-0.75mm}r_0),\Ext(x_{\cI_1},\hspace{-0.75mm}r_1))$ using $\Ext:\{0,\ldots,d-1\}^{n/\eta}\times\cR\rightarrow \sbin^\ell$, the $2$-universal hash function known from quantum key distribution~\cite{renato:diss}.
\item {\bf Bob: } Computes $c\in\sbin$ with $\cI=\cI_c$, and $x_{\cI}$ from $s$. He outputs $(c,y^\ell)\assign
(c,\Ext(s,r_c))$.
\end{protocol}

We provide only an overview of our proof since it closely follows the steps in~\cite{kww:nstorage}; details can be found in Appendix~\ref{sec:OT}. To show that the protocol is correct we first use Hoeffding's inequality~\cite{Hoeffding} to show that except with exponentially small probability $\exp(-2n/\eta)$, Bob learns a sufficient number of indices to retrieve the desired string $S_C$.

\subsection{Security against dishonest Alice}
To show that the protocol is secure against a cheating Alice, we have to show that there is no way for her to learn $C$, that is, which of the two strings
is known to honest Bob. We again provide an overview of our proof, and defer the complete technical details to Appendix~\ref{sec:OT}. 

First of all, note that the properties of weak string erasure ensure that a dishonest Alice does not know which $d$its~$x_{\cI}$ of $x^n$ are known to Bob, that is, she is ignorant about the index set~$\cI$. This is similar to the proof in~\cite{kww:nstorage}. However, for our new protocol we encounter an additional difficulty
since we need that even the truncated set $\cI_{\rm tr}$ is uniform over subsets of size $n/\eta$, but $\cI$ is no longer distributed uniformly over $2^{[n]}$. 
Formally, the probability of a given truncated set $\cI_{\rm tr}$ can be written in terms of the probability $p(\bar{A})$ that $|\cI| \geq n/\eta$, as follows:
\begin{equation}
	p(\cI_{\rm tr}|\bar{A}) = \sum_{\substack{\cI\\|\cI| \geq n/\eta}}\frac{p(\cI|\bar{A})}{\binom{|\cI|}{ n/\eta}} = \frac{1}{p(\bar{A})}\sum_{\cI}\frac{p(\cI)}{\binom{|\cI|}{n/\eta}}\ ,
\end{equation}
independent of the choice of truncation as desired. Here, $1/\binom{|\cI|}{n/\eta}$ is the probability of choosing the particular subset $\cI_{\rm tr}$ from $\cI$ and $p(\cI|\bar{A})$ is the conditional probability of a set $\cI$, given that $|\cI| \geq n/\eta$. The final equality is simply an application of Bayes' rule, $p(\bar{A})p(\cI|\bar{A}) = p(\bar{A}|\cI)p(\cI)$.
Finally, the fact that $\cI_{\rm tr}$ is uniform together with the properties of interactive hashing~\cite{savvides:diss} ensure that she cannot gain any information which of the two subsets $\mathcal{I}_0$ and ${\mathcal{I}_1}$ of bits are known to Bob. Hence, Alice cannot learn $C$ as desired.

\subsection{Security against dishonest Bob}
Again, it follows from weak string erasure that a dishonest Bob gains only a limited amount of information about the string $X^n$. The properties of interactive hashing ensure that Bob has very little control over the subset $\mathcal{I}_{1-c}$ chosen by the interactive hashing. Therefore, by the results on min-entropy sampling~\cite{kr:sampling}, Bob has only limited information about the $d$its in this subset. Privacy amplification~\cite{renato:diss,renato:compose} can then be used to turn this into almost complete ignorance.

\section{Conclusion}

We have shown that any two-party cryptographic primitive can be implemented securely in the setting of bounded quantum storage, even if the adversary can store all but a fraction of the transmitted pulses. 
This is optimal, in the sense that we could never hope to achieve security if the cheating party could store \emph{all} quantum communication made available to him. 
This demonstrates that there is no physical principle that prevents us from achieving security even with a very high storage rate $\nu < 1$. We have also shown in the noisy-storage setting that
security is possible for a much larger range of noisy quantum memories.

To achieve our result we use higher dimensional states which are very difficult to create in practice. It is therefore an interesting open question, whether the same result could 
be obtained using merely BB84 encoded qubits. Note, however, that our approach merely relies on the existence of entropic uncertainty relations for multiple encodings, and our protocols
and proofs are completely analogous if we were to use any other encodings for which strong uncertainty relations are known to exist.
For example, it is conceivable that uncertainty relations for multiple encodings can be based on top of BB84 encoded qubits~\cite{patrick:personal}, which would immediately lead to a 
protocol that is easy to implement experimentally. 

\acknowledgments
We thank Robert K{\"o}nig and J{\"u}rg Wullschleger for many interesting and useful discussions, as well 
as comments on an earlier draft. We also thank Christian Schaffner for comments.
PM and SW were supported by NSF grants PHY-04056720 and PHY-0803371. SW was supported by
the National Research Foundation and the Ministry of Education, Singapore. 
Part of this work was done while SW was at the Institute for Quantum Information, Caltech.

\appendix


\section{Weak String Erasure}\label{sec:wse}

To formally state our result, let us first define non-uniform weak string erasure.
This definition closely follows the one of~\cite{kww:nstorage}, except that the string $X^n$ is now chosen from a larger
alphabet and the indices in $\cI \subseteq [n]$ are not chosen uniformly at random.
Instead, the probability $p$ that honest Bob learns the value of $X_i$ for $i \in [n]$ is equal to the probability that 
he chooses the same basis as Alice, i.e., $p = 1/(d+1)$.
In the definition below, we will need to talk about distributions over subsets $\cI \subseteq [n]$. 
Clearly, the probability that Bob learns a particular subset $\cI$ satisfies
\begin{align}\label{eq:probDist}
	\Pr(\cI) = p^{|\cI|} (1-p)^{n-|\cI|}
\end{align}
Note that we can write the subset $\cI$ as a string $(y_1,\ldots,y_n) \in \{0,1\}^n$ where $y_i = 1$ if and only if
$i \in \cI$, allowing us to identify $\ket{\cI} := \ket{y_1} \otimes \ldots \otimes \ket{y_n}$. The probability
distribution over subsets $\cI \subseteq [n]$ can then be expressed as (see also~\cite{kww:nstorage})
\begin{align}
	\Psi(p) := \sum_{\cI \subseteq 2^{[n]}} p^{|\cI|} (1-p)^{n-|\cI|} \proj{\cI}\ .
\end{align}
Furthermore, we will follow the notation of~\cite{kww:nstorage} and use
\begin{align}
	\tau_{\cS} := \frac{1}{|\cS|} \sum_{s \in \cS} \proj{s}\ ,
\end{align}
to denote the uniform distribution over a set $\cS$. 

\begin{definition}[\textbf{Non-uniform WSE}]\label{def:wse}
An $(n,\lambda, \varepsilon,p,d)$--WSE scheme is a protocol between A and B satisfying the following properties:

\textbf{Correctness:} If both parties are honest, then there exists an ideal state $\sigma_{X^{n}\mathcal{I}X_{\mathcal{I}}}$ is defined such that:
\begin{enumerate}
\item The joint distribution of the $n$-dit string $X^{n}$ and subset $\mathcal{I}$ is given by
\begin{equation}\label{eq:wsecorrect}
\sigma_{X^{n}\cI} = \tau_{\{0,1,...,d-1\}^{n}}\otimes \Psi(p)\ ,
\end{equation}
\item The joint state $\rho_{AB}$ created by the real protocol is equal to the ideal state: $\rho_{AB} = \sigma_{X^{n}\cI X_{\cI}}$ where we identify $(A,B)$ with $(X^{n},\cI X_{\cI})$.
\end{enumerate}

\textbf{Security for Alice:} If A is honest, then there exists an ideal state $\sigma_{X^{n}B'}$ such that 
\begin{enumerate}
\item The amount of information $B'$ gives Bob about $X^{n}$ is limited:
\begin{equation}\label{eq:honestA}
\frac{1}{n}H_{\infty}(X^{n}|B')_{\sigma} \geq \lambda
\end{equation}
\item The joint state $\rho_{AB'}$ created by the real protocol is $\eps$-close to the ideal state, i.e. $\sigma_{X^{n}B'} \approx_{\varepsilon} \rho_{AB'}$ where we identify $(X^{n},B')$ with $(A,B')$.
\end{enumerate}

\textbf{Security for Bob:} If B is honest, then there exists ideal state $\sigma_{A'\hat{X}^{n}\cI}$ where $\hat{X}^{n} \in \{0,1,...,d-1\}^{n}$ and $\cI \subseteq [n]$ such that
\begin{enumerate}
	\item The random variable $\cI$ is independent of $A'\hat{X}^{n}$ and distributed over $2^{[n]}$ according to the probability distribution given by~\eqref{eq:probDist}:
\begin{equation}\label{eq:honestB}
\sigma_{A'\hat{X}^{n}\cI} = \sigma_{A'\hat{X}^{n}} \otimes \Psi(p)\ .
\end{equation}
\item The joint state $\rho_{A'B}$ created by the real protocol is equal to the ideal state: $\rho_{A'B} = \sigma_{A'(\cI\hat{X}_{\cI})}$, where we identify $(A',B)$ with $(A',\cI\hat{X}_{\cI})$.
\end{enumerate}
\end{definition}

We are now ready to state our result for non-uniform weak string erasure more formally. We first state the general result for quantum memories, and then focus on the tensor-product channels of the type $\cF = \cN^{\otimes \nu \cdot n}$. 

\begin{theorem}\label{thm:appwse}
\begin{enumerate}
\item[(i)] Let $\delta\in ]0,\frac{1}{2}[$ and let Bob's storage be given by $\cF: \cB(\hin) \rightarrow \cB(\hout)$. Then Protocol~1 is an $(n,\lambda(\delta,d),\eps(\delta,d),1/(d+1),d)$--WSE protocol with
min-entropy rate
\begin{align*}
\lambda(\delta,d)=-\lim_{n\rightarrow \infty}\frac{1}{n} P^\cF_{succ}\left(\left(\log(d+1)-1-\delta\right)\cdot n\right)\ ,
\end{align*}
and error $\eps(\delta,d) = 2\exp(-f(\delta,d) n)$ with
\begin{align}
	f(\delta,d) := \frac{(\delta/4)^{2}}{32\left(\log((d+1)\cdot d)+\log\frac{4}{\delta}\right)^2} > 0.\label{eq:thmFdef}
\end{align}
\item[(ii)] Suppose $\cF=\cN^{\otimes \nu n}$ for a storage rate~$\nu>0$, $\cN$ satisfying the \scp\ and having capacity~$C_\cN$ bounded by
\begin{align*}
C_\cN\cdot\nu<\log(d+1)-1\ .
\end{align*} 
Let $\delta\in ]0,\frac{1}{2}-C_\cN\cdot \nu[$.
Then Protocol~1 is an $(n,\tilde{\lambda}(\delta,d),\eps(\delta,d),1/(d+1),d)$--WSE protocol 
for sufficiently large~$n$, where 
\begin{align*}
\tilde{\lambda}(\delta,d) &=\nu \cdot \gamma^\cN\left(\frac{\log(d+1)-1-\delta}{\nu}\right)\ .
\end{align*}
\end{enumerate}
\end{theorem}

Note that when $\mathcal{N} = \id_d$ then $C_{\mathcal{N}} = \log d$, so that the bound in~\eqref{eq:security} holds for a storage rate of 
\[
\nu < \frac{\log(d+1) -1}{\log d} \approx 1, \; \textrm{for large} \; d.
\] 
Thus for the case of bounded storage, security can in principle be obtained for any storage rate $\nu < 1$, provided we choose a large enough system size $d$.

\subsection{Security for honest Alice}\label{sec:wseA}

Let us now first consider the case of dishonest Bob.
The main difference from~\cite{kww:nstorage} in proving security lies in the use of the uncertainty relation for the full set
of $d+1$ mutually unbiased bases in prime power dimensions~\cite{sanchez:entropy}.
To see where we will make use of this relation, note that analogous to~\cite{kww:nstorage} we can model Bob's attack as a CPTP map 
$\cE: \cB((\mathbb{C}^{d})^{\otimes n}) \rightarrow \cB(\mathcal{H}_{\rm in}\otimes \cH_{K})$. Then, for any input state $\rho \in (\mathbb{C}^{d})^{\otimes n}$ provided by Alice before the waiting time, 
he obtains an output state $\zeta_{Q_{\rm in} K} = \cE(\rho)$, where $Q_{\rm in}$ is the quantum information he puts into his quantum storage and $K$ is any additional classical information he retains. 
Hence, the joint state of Alice and Bob before his storage noise is applied is of the form
\begin{align}
&\rho_{X^{n}\Theta^{n}KQ_{in}} = \frac{1}{d^{n}(d+1)^{n}}\sum_{x^{n}, \theta^{n}, k} P_{K|X^{n}=x^{n},\Theta^{n}=\theta^{n}}(k) \nonumber\\
&\qquad\underbrace{\ket{x^{n}}\bra{x^{n}}\otimes\ket{\theta^{n}}\bra{\theta^{n}}}_{\textrm{Alice}}\otimes \underbrace{\ket{k}\bra{k}\otimes\zeta_{x^{n}\theta^{n} k}}_{\textrm{Bob}}
\end{align}
where $\zeta_{x^{n}\theta^{n} k}$ is the state on $\hin$ depending on Alice's choice of string $x^n$, bases $\theta^n$ and Bob's classical information $k$.
Bob's storage then undergoes noise described by $\mathcal{F}: \mathcal{B}(\hin) \rightarrow \cB(\hout)$, and the state evolves to $\rho_{X^{n}\Theta^{n}K\mathcal{F}(Q_{\rm in})}$. 
After time $\Delta t$ Bob also receives the basis info $\Theta^{n} = \theta^{n}$. Then their joint state is
\begin{align}
&\rho_{X^{n}\Theta^{n}K\mathcal{F}(Q_{\rm in})} = \frac{1}{d^{n}(d+1)^{n}}\sum_{x^{n},\theta^{n},k} P_{K|X^{n}=x^{n},\Theta^{n}=\theta^{n}}(k)\nonumber \\
&\qquad\underbrace{\ket{x^{n}}\bra{x^{n}}}_{\textrm{Alice}}\otimes\underbrace{\ket{\theta^{n}}\bra{\theta^{n}}\otimes\mathcal{F}(\zeta_{x^{n}\theta^{n}k})}_{\textrm{Bob B'}}
\end{align}
where Bob now holds $B' = \Theta^{n}K\mathcal{F}(Q_{\rm in})$.

Our goal is to show
that for any cheating strategy of dishonest Bob, his min-entropy about the string $X^{n} = (X_{1},...,X_{n})$ is large, using an entropic uncertainty relation. Recall that the set of $(d+1)$ MUBs in $\mathbb{C}_{d}$ satisfies~\cite{sanchez:entropy} (see~\cite{ww:ursurvey} for a simple proof)
\begin{equation}\label{eq:eur}
\frac{1}{d+1}\sum_{i=1}^{d+1}H(\cB_{i}|\rho) \geq \log(d+1) - 1, \,\forall \,\rho\,\in\,\mathbb{C}_{d},
\end{equation}
where 
\begin{align}
H(\cB_{i}|\rho) = -\sum_{x}\textrm{Tr}(\ket{b_{i}^{x}}\bra{b_{i}^{x}}\rho)\log\textrm{Tr}(\ket{b_{i}^{x}}\bra{b_{i}^{x}}\rho)
\end{align}
is the Shannon entropy of the probability distribution
induced by measuring the state $\rho$ in the basis $\cB_{i}$. 
This lower bound, along with the uncertainty relation for the smooth min-entropy from~\cite[Theorem 4.22]{chris:diss} gives
\begin{equation}
H^{\varepsilon/2}_{\infty}(X^{n}|\Theta^{n})_{\rho} \geq \left(\log(d+1)-1 - \frac{\delta}{2}\right)n
\end{equation}
for any $0<\delta<\frac{1}{2}$ with
\begin{equation}
\varepsilon = 2\exp\left(-\frac{(\delta/4)^{2}n}{32\left(\log((d+1)\cdot d)+ \log\frac{4}{\delta}\right)^2}\right).
\end{equation}
Finally, we make use of Lemma~2.2 in~\cite{kww:nstorage} that relates the smooth min-entropy to the maximal decoding probability, $P^{\cF}_{succ}$, to get,
\begin{eqnarray}
&H^{\varepsilon}_{\infty}(X^{n}|\Theta^{n}K\mathcal{F}(Q_{in}))_{\rho} \nonumber \\
&\qquad \geq - \log P^{\mathcal{F}}_{succ}\left(n\left(\log(d+1)-1 - \frac{\delta}{2}\right) -\log\frac{2}{\varepsilon}\right) \nonumber \\
&\qquad \geq -\log P^{\mathcal{F}}_{succ}\left(n(\log(d+1)-1)-n\frac{\delta}{2}\right) \nonumber
\end{eqnarray}
where the second inequality follows from the monotonicity of $P_{\rm succ}^{\cF}$ and the fact that $\log\frac{2}{\varepsilon} < \frac{\delta}{2}n$ for $0<\delta<\frac{1}{2}$. By definition of the smooth min-entropy, this implies that there exists
an ideal state $\sigma_{X^{n}B'}$ such that
\begin{enumerate}
\item $\sigma_{X^{n}B'} \approx_{\varepsilon} \rho_{X^{n}B'}$,
\item $\frac{1}{n}H_{\infty}(X^{n}|B')_{\sigma} \geq -\frac{1}{n}\log P^{\mathcal{F}}_{succ}\left(n\log(d+1)-n - \frac{\delta}{2}n\right)$,
\end{enumerate}
which proves part (i) of Theorem~\ref{thm:appwse}.

In the special case that $\mathcal{F}$ is the tensor product channel $\mathcal{F} = \mathcal{N}^{\otimes\nu n}$, where $\mathcal{N}$ satisfies the strong converse property and 
$C_{\mathcal{N}}\cdot \nu < \log(d+1) -1$, following the same 
steps as in~\cite{kww:nstorage} we obtain that there
exists an ideal state $\sigma_{X^{n}B'}$ that is $\varepsilon$-close to $\rho_{X^{n}B'}$ and has a min-entropy
\begin{equation}\label{eq:security}
\frac{1}{n}H_{\infty}(X^{n}|B')_{\sigma} \geq \nu.\gamma^{\mathcal{N}}\left(\frac{\log(d+1)-1-\frac{\delta}{2}}{\nu}\right) > 0\ ,
\end{equation}
where $\gamma^{\mathcal{N}}(\cdot)$ is the strong converse parameter of the channel $\mathcal{N}$~\cite{rs:converse}. This proves part (ii) of Theorem~\ref{thm:appwse}.

\subsection{Security for honest Bob}\label{sec:wseB}

The proof of security when Alice is dishonest is essentially analogous to~\cite{kww:nstorage} (see Section 3.4 and Figures 7 and 8), where we introduce an imaginary ``simulator'' with perfect quantum memory to define the 
desired ideal state. We hence merely state how to adapt the proof of~\cite{kww:nstorage}: here we naturally obtain $\Psi(p)$ in place of the uniform distribution $\tau_{2^{[n]}}$ in our simulation. Similarly, the subset $\cI$ is not chosen uniformly
at random, but with probability
\begin{align}
	\Pr(\cI) := \left(\frac{1}{d+1}\right)^{|\cI|} \left(\frac{d}{d+1}\right)^{n-|\cI|}\ .
\end{align}

\section{Oblivious Transfer from Weak String Erasure}\label{sec:OT}

We are now ready to show how oblivious transfer can be obtained even from the non-uniform variant of weak string erasure.
To formally state our result we begin with the definition of oblivious transfer from~\cite{kww:nstorage}. 

\begin{definition}
An {\em $(\ell,\eps)$--fully randomized oblivious transfer (FROT) scheme} is a protocol between Alice and Bob satisfying the following:
\begin{description}
\item[Correctness:] If both parties are honest, then the ideal state
$\sigma_{S_0^\ell S_1^\ell C S_C^\ell}$ is defined such that
\begin{enumerate}
\item The distribution over $S_0^\ell$, $S_1^\ell$ and $C$ is uniform:
$$
\sigma_{S_0^\ell S_1^\ell C}=\tau_{\{0,1\}^\ell} \otimes
\tau_{\{0,1\}^\ell} \otimes \tau_{\{0,1\}}\ .
$$
\item The real state $\rho_{S_0^\ell S_1^\ell C Y^\ell}$ created
during the protocol is $\eps$-close to the ideal state:
\begin{align}
\rho_{S_0^\ell S_1^\ell C Y^\ell} \approx_\eps \sigma_{S_0^\ell
S_1^\ell C S_C^\ell}\ ,
\end{align}
where we identify $A=(S_0^\ell,S_1^\ell)$ and $B=(C,Y^\ell)$.
\end{enumerate}
\item[Security for Alice:] If Alice is honest, then there exists an
ideal state $\sigma_{S_0^\ell S_1^\ell B'C}$,
where $C$ is a random variable on $\{0,1\}$, such that
\begin{enumerate}
\item Bob is ignorant about $S_{1-C}^\ell$:
$$
\sigma_{S_{1-C}^\ell S_C^\ell B'C} \approx_\eps
\tau_{\{0,1\}^\ell}\otimes\sigma_{S_C^\ell B'C}\ .
$$
\item The real state $\rho_{S_0^\ell S_1^\ell B'}$ created during the
protocol is $\eps$-close to the ideal state:
$$
\rho_{S_0^\ell S_1^\ell B'} \approx_\eps \sigma_{S_0^\ell S_1^\ell B'}\ .
$$
\end{enumerate}

\item[Security for Bob:] If Bob is honest, then there exists an ideal state $\sigma_{A'S_0^\ell S_1^\ell C}$ such that
\begin{enumerate}
\item Alice is ignorant about $C$:
$$
\sigma_{A'S_0^\ell S_1^\ell C} = \sigma_{A'S_0^\ell
S_1^\ell}\otimes\tau_{\{0,1\}}\ .
$$
\item The real state $\rho_{A'C Y^\ell}$ created during the protocol is $\eps$-close to the ideal state:
$$
\rho_{A'C Y^\ell} \approx_\eps \sigma_{A'CS_C^\ell}\ ,
$$
where we identify $B = (C,Y^\ell)$.
\end{enumerate}
\end{description}
\end{definition}

In the main part of this text, we had restricted ourselves to considering a simplified protocol containing all the essential ideas of the protocol below.
The actual protocol is very similar, but for technical reasons we will work with $m$ blocks of $\beta$ elements each, instead of sampling individual elements $X_j$.
Fortunately, the protocol we will use for the case of non-uniform weak string erasure remains the same as in the case of weak string erasure with a small modification.
Since $p \neq 1/2$, the expected number of $d$its $X_j$ that Bob will learn is of course no longer roughly $n/2$ as in the original setting~\cite{kww:nstorage}. 
This requires the introduction of a new parameter $\eta$ such that with high probability Bob will learn $n/\eta$ of the string's entries.
We again require an encoding of subsets as strings. Since our
subsets will now be smaller, we choose $t$ such that $2^t\leq \binom{m}{m/\eta}\leq 2\cdot 2^t$, and an injective encoding $\Enc:\sbin^t\rightarrow\cT$,
where $\cT$ is the set of possible subsets of size $m/\eta$.
Note that this again means that not all subsets can be encoded but at least half of them will.

\begin{protocol}{2}{WSE-to-FROT}{
Parameters: Set $\eta := 2(d+1)$. Integers $n,\beta$ such that  $m:=n/\beta$ 
is a multiple of~$\eta$.  
Outputs: $(s_0^\ell,s_1^\ell) \in
\{0,1\}^{\ell} \times \{0,1\}^\ell$ to Alice, and $(c,y^\ell) \in \{0,1\} \times \{0,1\}^\ell$
to Bob.}
\item {\bf Alice and Bob: }  Execute $[n,\lambda,\eps,1/(d+1),d]$--WSE.
	Alice gets a string $x^n \in \{0,1,\ldots,d-1\}^n$, Bob a set $\cI \subset [n]$ and a string $s = x_\cI$. If $|\cI| < n/\eta$, then Bob simply chooses $\cI_{\rm tr}$ from all subsets of size $|\cI|  = n/\eta$ uniformly at random. 
Otherwise, he randomly truncates $\cI$ to $\cI_{\rm tr}$ of size $n/\eta$, and deletes the corresponding values in $s$.

We arrange $x^n$ into a matrix ${\bf z} \in \Mat_{m \times \beta}(\{0,1,\ldots,d-1\})$, by ${\bf z}_{j,\alpha} := x_{(j-1) \cdot \beta + \alpha}$ for
$(j,\alpha) \in [m] \times [\beta]$.
 
\item {\bf Bob: }
\begin{enumerate}
\item Randomly chooses a string $w^t\in_R \{0,1\}^t$ corresponding to an encoding  of a subset $\Enc(w^t)$ of $[m]$ with $m/\eta$ elements. 
\item Randomly partitions the $n$~$d$its of $x^n$ into $m$ blocks of $\beta$~$d$its each: 
He randomly chooses a permutation $\pi:[m]\times[\beta]\rightarrow [m]\times [\beta]$ of the entries of ${\bf z}$ 
such that he knows $\pi({\bf z})_{\Enc(w^t)}$ (that is, these $d$its are permutation of the $d$its of~$s$). Formally, $\pi$ is uniform over permutations satisfying the following condition: for all $(j, \alpha) \in [m] \times [\beta]$ and $(j', \alpha') := \pi (j,\alpha)$, we have $(j-1) \cdot \beta + \alpha \in \cI  \Leftrightarrow j' \in \Enc(w^t)$.

\item Bob sends $\pi$ to Alice.
\end{enumerate}
\item {\bf Alice and Bob:} Execute interactive hashing with Bob's input equal to $w^t$. They obtain $w_0^t,w_1^t \in \{0,1\}^t$ with $w^t\in\{w_0^t,w_1^t\}$. 
\item {\bf Alice: } Chooses $r_{0}, r_{1} \in_R \cR$ and sends them to Bob.
\item {\bf Alice: } Outputs $(s_0^\ell,s_1^\ell) \assign (\Ext(\pi({\bf z})_{\Enc(w_0^t)},\hspace{-0.75mm}r_0),
\Ext(\pi({\bf z})_{\Enc(w_1^t)},\hspace{-0.75mm}r_1))$.
\item {\bf Bob: } Computes $c$, where $w^t=w_c^t$, and $\pi({\bf z})_{\Enc(w^t)}$ from $s$. He outputs $(c,y^\ell) \assign (c,\Ext(\pi({\bf z})_{\Enc(w^t)},\hspace{-0.75mm}r_c))$.
\end{protocol}

\begin{theorem}[Oblivious Transfer]
For any $\omega \geq (d+1)$ and $\beta \geq \max\{67,256 \omega^2/\lambda^2\}$, the protocol WSE-to-FROT implements an $(\ell,43 \cdot 2^{-\frac{\lambda^2}{512 \omega^2 \beta} n} + 2 \eps)$--FROT from  one instance of of $[n, \lambda,
\eps,p,d]$--non-uniform WSE, where
$\ell := \left \lfloor \left(\left(\frac{\omega-1}{\omega}\right)\frac{\lambda}{4(d+1)}-\frac{\lambda^2}{512 \omega^2\beta}\right)n - \frac{1}{2} \right \rfloor$.
\end{theorem}

\subsection{Security for Bob}

We first show that the protocol is secure against a cheating Alice. This can again be done following the steps of~\cite{kww:nstorage} taking the non-uniformity into account.
Formally, let $\tilde\rho_{A'' C Y^\ell }$ denote the joint state at the end of the protocol, 
where $A''$ is the quantum output of a malicious Alice and $(C,Y^\ell)$ is the classical output of an honest Bob. Following the same steps as in~\cite{kww:nstorage} we can
construct an ideal state $\tilde \sigma_{A'' W_0^\ell W_1^\ell C}
= \tilde \sigma_{A'' W_0^\ell W_1^\ell} \otimes \tau_{\{0,1\}}$ that satisfies
$\tilde \rho_{A'' C Y^\ell } = \tilde \sigma_{A'' C W_C^\ell } = \tilde{\sigma}_{A'' C S_C^\ell}$.

It now again remains to be shown that Alice does not learn anything about $C$, that is, $\tilde \sigma_{A'' S_0^\ell S_1^\ell C} = \tilde \sigma_{A'' S_0^\ell S_1^\ell} \otimes \tau_{\{0,1\}}$. 
From the properties of non-uniform WSE it follows that $\sigma_{A'\hat{X}^n\cI}=\sigma_{A'\hat{X}^n}\otimes\Psi(1/(d+1))$. Since Bob randomly truncates $\cI$ to $\cI_{\rm tr}$ such that $| \cI_{\rm tr} | = n/\eta$, the truncated set is independent of $A'$. Furthermore, although $\cI$ is not distributed uniformly over $2^{[n]}$, we can show that the truncated set $\cI_{\rm tr}$ is indeed distributed uniformly over all subsets of size $n/\eta$. Intuitively this follows from the fact that the distribution of the set $\cI$ depends only on $|\cI|$, the number of elements in $\cI$. Formally, the probability of a given truncated set $\cI_{\rm tr}$ can be written in terms of the probability $p(\bar{A})$ that $|\cI| \geq n/\eta$ as follows
\begin{eqnarray}
	p(\cI_{\rm tr} | \bar{A}) &=& \sum_{\substack{\cI \subseteq [n]\\|\cI| \geq n/\eta}}\frac{p(\cI|\bar{A})}{\binom{|\cI|}{ n/\eta}}p(|\cI| \geq n/\eta) \\
&=&  \frac{1}{p(\bar{A})}\sum_{\cI}\frac{p(\cI)}{\binom{|\cI|}{n/\eta}}, \nonumber
\end{eqnarray}
independent of the choice of truncation. Here $1/\binom{|\cI|}{n/\eta}$ is the probability that we pick a particular $\cI_{\rm tr}$ from the original $\cI$ 
and $p(\cI|\bar{A})$ is the conditional probability of a set $\cI$, given that Bob obtains a sufficient number of indices. The last step is simply an application of Bayes' rule, $p(\bar{A})p(\cI|\bar{A}) = p(\bar{A}|\cI)p(\cI)$
where $p(\bar{A}|\cI) = 1$ for the subsets $\cI$ in the sum.
Note that if $|\cI| < n/\eta$ then Bob chooses a subset of the desired size uniformly at random from all subsets of size $|\cI| = n/\eta$ and hence $\Pr(\cI_{\rm tr})$ is always uniform.
Hence, conditioned on any fixed $W^t = w^t$, the permutation $\Pi$ is uniform and independent of $A'$. It follows that the string $W^t$ is also uniform and independent of $A'$ and $\Pi$. From the properties of interactive hashing we are guaranteed that  $C$ is uniform and independent of Alice's view afterwards, and hence,
 \[\tilde \sigma_{A'' S_0^\ell S_1^\ell C} = \tilde \sigma_{A''
S_0^\ell S_1^\ell} \otimes \tau_{\{0,1\}}\;.\]

\subsection{Security for Alice}

The security proof for the case that Bob is dishonest is analogous to~\cite{kww:nstorage}, this time employing~\cite[Lemma 2.5]{kww:nstorage} with a subset size of $|\cS| = m/\eta$.

\subsection{Correctness}

It remains to prove that if both parties are honest, then honest Bob can indeed learn the desired $S_C$.
This requires us to show that for our choice of $\eta$, Bob can learn sufficiently many indices $i \in [n]$. 

\begin{lemma}[Correctness] 
\label{lem:FROT-correctness}
Protocol WSE-to-FROT satisfies correctness with an error of
\[ 43 \cdot 2^{-\frac{\lambda^2}{512 \omega^2\beta} n}\;.\]
\end{lemma}

First we show using the Hoeffding bound~\cite{Hoeffding}, that the probability that a subset of $[n]$ where each entry is chosen with probability $p=1/(d+1)$
has less than $n/\eta$ elements is at most $\exp(-2n/\eta^2)$. Consider a sequence of independent random variables $\{X_{1}, \ldots,X_{n}\}$, which are bounded as follows: $\textrm{Pr}(X_i - {\rm\textbf{E}}(X_i) \in [a_i,b_i]) = 1, \; \forall 1<i<n$. Then, Hoeffding's inequality states that the sum $S = X_1 + \ldots + X_n$ satisfies,
\begin{align}
{\rm Pr}({\rm \textbf{E}}(S) - S \geq t) &\leq \exp\left(-\frac{2t^2}{\sum_{i=1}^{n}(b_i - a_i)^2}\right)
\end{align}

In our context, $X_{i}$ is the binary variable which takes on the value $1$ if the index $i \in \cI$, and $0$ otherwise. The sum $S$ is thus simply equal to $|\cI|$, the number of elements in the index set $\cI$, which is a random subset of $[n]$. 
For the case of $d+1$ encodings, ${\rm Pr}(X_i = 1) = 1/(d+1)$ and ${\rm Pr}(X_i = 0) = d/(d+1)$, so that the expectation value satisfies
\begin{align}
	{\rm \textbf{E}}(S)  = {\rm \textbf{E}}(|\cI|) = \frac{n}{d+1}\ .
\end{align}
Furthermore, we can take $a_i = 0$ and $b_i = 1$ for all $i$.
Applying Hoeffding's inequality to the sum $S = |\cI|$ gives
\begin{align}
&{\rm Pr}(\frac{n}{d+1} - |\cI| \geq \frac{n}{d+1} - \frac{n}{\eta}) \leq\nonumber\\
&\qquad\exp\left(-2n\left[\frac{1}{d+1}-\frac{1}{\eta}\right]^2\right).
\end{align}
Rearranging terms, we obtain the probability that a random set $\cI$ has less than $n/\eta$ elements:
\begin{align}
&{\rm Pr}(|\cI| \leq \frac{n}{\eta}) \leq \exp\left(-2n\left[\frac{1}{d+1}-\frac{1}{\eta}\right]^2\right).
\end{align}
Since our work is mainly a proof of principle, we do not yet care about optimality or efficiency. We simply pick a choice of $\eta$ that will satisfy this condition, and set $\eta = 2(d+1)$. Thus, the probability that a random subset of $[n]$ has less than $n/\eta$ elements is at most $\exp(-2n/\eta^2)$.

Let $\xi := 2^{-n/\eta^2}$. We have to show that the state $\tilde \rho_{S_0^\ell S_1^\ell C Y^\ell}$ at the end of the protocol is close to the given ideal state $\tilde \sigma_{S_0^\ell S_1^\ell C S^\ell_C}$. As shown above, the probability that a subset of $[n]$ 
has less than $n/\eta$ elements is at most 
\begin{align}
	\exp(-2n/\eta^2) \leq \xi\ . 
\end{align}
Hence, the probability that Bob does not learn sufficiently many indices when both parties are honest is at most $\xi$. Let $\cA$ be the event that $|\cI| \geq n/\eta$. 
It remains to show that the state $\tilde \rho_{S_0^\ell S_1^\ell C Y^\ell \mid \cA}$ is close to the given ideal state $\sigma_{S_0^\ell S_1^\ell C S^\ell_C}$. 

Note that the correctness condition of WSE ensures that the state created by WSE is equal to $\rho_{X^n \cI X_{\cI}} = \sigma_{X^n \cI X_{\cI}}$, where $\sigma_{X^n \cI} = \tau_{\{0,1,...,d-1\}^{n}}\otimes \Psi(1/(d+1))$. 
Since $\cI_0$ and $\cI_1$ are chosen independently of $X^n$, $X_{\cI_0}$ and $X_{\cI_1}$ have a min-entropy of $n/\eta$ each. Since $\ell \leq n/2\eta \leq n/\eta -
2 \log 1/\xi$, it follows from privacy amplification that $S^\ell_C$ 
is independent and $\xi$-close to uniform.
Since dishonest Bob is only more
powerful than honest Bob, we furthermore have from the proof against dishonest Bob that $S^\ell_{1-C}$
is independent and uniform except with an error of at most
$\hat{\eps} = 41 \cdot 2^{-\frac{\lambda^2}{512 \omega^2 \beta} n}$, where we used the fact that
Bob is also honest during weak string erasure ($\eps = 0$).
Finally, by the same arguments showing security for Bob we have that $C$ is uniform
and independent of $S^\ell_0$ and $S^\ell_1$. Hence,
\[\rho_{S_0^\ell S_1^\ell C \mid \cA} \approx_{\xi + \hat{\eps}} \sigma_{S_0^\ell
S_1^\ell C}\;.\]
Since the extra condition on the permutation $\Pi$ implies that Bob
can indeed calculate
$\Pi({\bf Z})_{\Enc(W)}$ from $X_{\cI}$, we have that
$Y^\ell = S_C^\ell$. Using $\Pr[\cA] \geq 1 - \xi$, we get
\[\rho_{S_0^\ell S_1^\ell C Y^\ell} \approx_{2 \xi + \hat{\eps}} \sigma_{S_0^\ell
S_1^\ell C S^\ell_C}\;.\]
Finally, $\lambda \leq 1$, $\beta > 1$ and $\omega\geq (d+1)$ give us $1/\eta^{2} = 1/(4(d+1)^{2}) > \lambda^2/(512\omega^{2}\beta)$.
Adding up all errors and noting that
\[2 \cdot 2^{-\frac{1}{\eta^2}n} \leq 2 \cdot 2^{-\frac{\lambda^2}{512 \omega^2\beta} n}\;,\]
gives our claim.

\end{document}